\documentclass[twocolumn,english,prb,showpacs,superscriptaddress]{revtex4-1}
\usepackage{hyperref}
\usepackage{amssymb}
\usepackage{graphicx}
\usepackage{amsmath}
\usepackage{mathrsfs}
\usepackage[usenames]{color}
\usepackage{wasysym}
\usepackage{bbold}
\usepackage{multirow}

\begin{document}

\title{Thermodynamics of the Hubbard model on  stacked honeycomb and square lattices}
\author{Jakub Imri\v{s}ka}
\affiliation{Theoretische Physik, ETH Zurich, 8093 Zurich, Switzerland}
\author{Emanuel Gull}
\affiliation{Department of Physics, University of Michigan, Ann Arbor, MI 48109, USA}
\author{Matthias Troyer}
\affiliation{Theoretische Physik, ETH Zurich, 8093 Zurich, Switzerland}

\begin{abstract}
We present a numerical study of the Hubbard model on simply stacked honeycomb and square lattices, motivated by a recent experimental realization of such models with ultracold atoms in optical lattices. We perform simulations with different interlayer coupling and interaction strengths and obtain N\'eel transition temperatures and entropies.
We provide data for the equation of state to enable comparisons of experiments and theory. We find an enhancement of the short-range correlations in the anisotropic lattices compared to the isotropic cubic lattice, in parameter regimes suitable for the interaction driven adiabatic cooling.
\end{abstract}


\maketitle

\section{Introduction}
The single-orbital Hubbard model, originally introduced to describe correlation driven metal-insulator transitions,\cite{Hubbard63} has been the subject of intensive study in recent years, as it is widely believed that its realization on a two-dimensional square lattice captures many of the salient features of high-temperature superconductivity.\cite{Scalapino94,Scalapino07} Apart from a Fermi liquid phase at weak interaction and large doping strength and a correlation driven insulating phase at half filling and large interaction strength, superconducting phases of various types,\cite{Zanchi96,Sorella02,Raghu10,Maier05_dwave} pseudogap behavior in the absence of long-range order,\cite{Huscroft01,Kyung06,Lin10,Gull13} ferromagnetic,\cite{Becca01,Nagaoka66,Chang10} and antiferromagnetic\cite{Staudt00} phases, as well as different types of stripe phases\cite{Steve00,Scuseria14}  have been proposed. 

Theoretical and numerical studies of the low temperature properties of the Hubbard model have proven to be difficult, especially in the strongly correlated regime where the interaction strength is comparable to the bandwidth and many low-lying degrees of freedom compete. Experimental realizations using cold atomic gas systems,\cite{Lewenstein:2007,Esslinger:2010ex} on various lattices in two and three dimensions, offer  an alternative route to increase our understanding the physics of this model. While the temperatures accessible by these experiment are still far above the superconducting phase transitions, a range of phenomena, including long range antiferromagnetic order in three dimensions, may soon be accessible.\cite{CoolingToAF:2011,Hulet:2014,ShortRangeCorrelationsExp,AnisotropicHubbard14,AntiFerromagneticCorrelations_Esslinger2015}

One of the  current challenges is the calibration of the precise parameters of experiments using ultracold atomic gases, and in particular their temperature or entropy. Numerical simulations of the model for a range of parameters have proven to be useful in this context, and especially quantities that show a strong dependence on temperature and are accessible both in simulation and experiment. An example are nearest neighbor the spin-correlations.\cite{ShortRangeCorrelationsExp} Comparison to numerics was able to  identify unexpected heating effects and could pinpoint the temperature down to which the experimental realization of the model was accurate.\cite{AnisotropicHubbard14}

Motivated by the physics of graphene and by the search for a spin liquid state at low temperature,\cite{SpinLiquid2010,NoSpinLiquid2012,AssaadPinningTheOrder2013} experimental realizations of the model on a honeycomb geometry have appeared~\cite{ArtificialGraphene13} and provided results in agreement with numerical calculations of the $2d$ model.\cite{ThermodynamicsOfHoneycomb2012,ThermodynamicsOfHoneycomb2013}
Complementary to studies on isotropic lattices, anisotropic lattices of various types, e.g. with couplings in the vertical axis chosen differently from in-plane couplings, can be realized.\cite{ShortRangeCorrelationsExp,ArtificialGraphene13} These models offer the possibility of studying a dimensional crossover between three, two and one dimensions and with this the possibility of tuning phase transitions to a more readily accessible regime. 

From the experimental perspective, layered systems are a natural setup to investigate quasi-2d physics. The reduced dimensionality may give rise to interesting phenomena, but the presence of the third dimension will affect some of the low temperature properties -- {\em e.g.} allowing for long range order at non-zero temperature which is absent in systems with continuous symmetries in two dimensions.\cite{MerminWagnerTheorem1966,Hohenberg1967} 

For the purpose of quantitative comparisons to cold atoms experiments, numerical simulations need to provide results at comparatively high temperature. For much of the parameter regime accessible to experiment,s high temperature series expansion and numerical linked cluster expansions seem to be sufficient. As the temperature is lowered outside of the convergence radius of these series, non-perturbative techniques are required. Cluster dynamical mean field methods in particular\cite{QClusterTheoriesReview2005}  are able to reach lower temperature in the thermodynamic limit both at and away from half filling and have been shown to be a reliable tool for this task.\cite{Fuchs:2011ch}

\begin{figure}[t]
\centering
\includegraphics[width=\columnwidth]{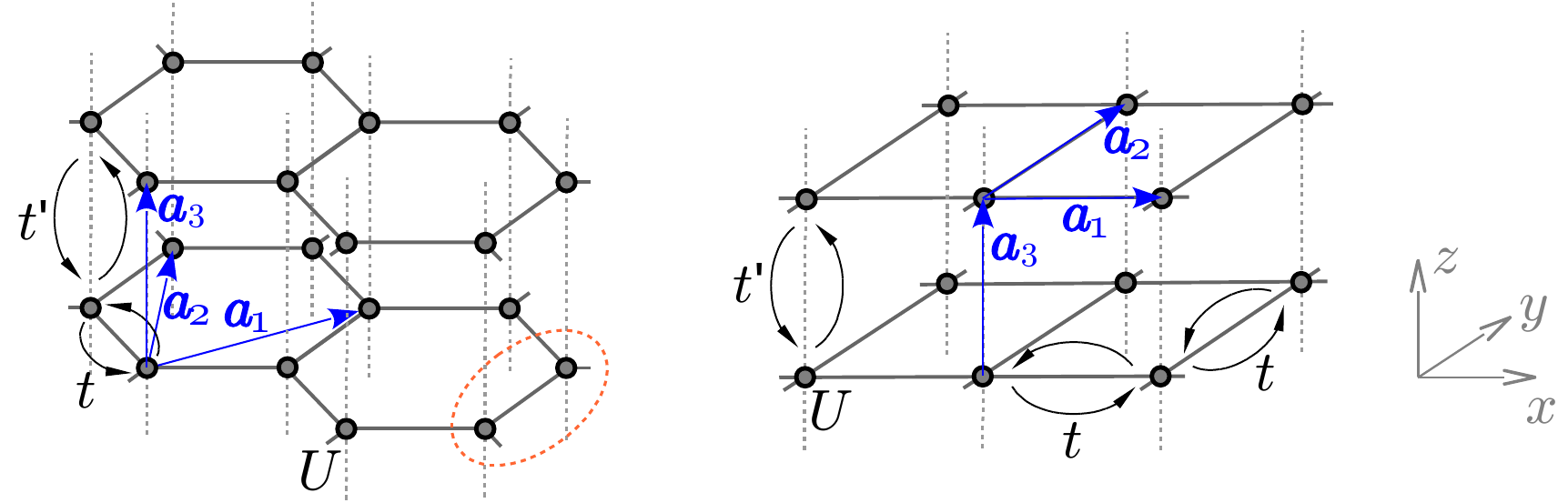}
\caption{Left panel: simply stacked honeycomb lattice with interaction $U$, in-plane hopping $t$, and inter-plane hopping $t^\prime$. The dashed ellipse denotes the unit cell, containing two sites. The lattice basis vectors ${\bf a}_i$ are shown in blue. Right panel: stacked square lattice. Here, the unit cell consists of a single site.}
\label{fig:lattices}
\end{figure}

Here we use these methods to study weakly to moderately coupled stacked honeycomb and square lattices, as depicted in Fig.~\ref{fig:lattices}. The Hamiltonian of the Hubbard model on these lattices is
\begin{eqnarray}
	\hat{H}&=& -t \sum_{\left\langle i,j \right\rangle,\sigma} \hat{c}^\dag_{i\sigma} \hat{c}_{j\sigma}
	           -t^\prime \sum_{\left\langle i,j \right\rangle^{\prime},\sigma} \hat{c}^\dag_{i\sigma} \hat{c}_{j\sigma} \nonumber \\
	       &&  -\mu \sum_{i,\sigma} \hat{n}_{i\sigma} %\nonumber \\
  +U \sum_{i} \hat{n}_{i\uparrow} \hat{n}_{i\downarrow}   ,
	 \label {eq:Ham}
\end{eqnarray}
where $\hat{c}^\dag_{i\sigma}$ ($\hat{c}_{i\sigma}$) creates (annihilates) a fermion at site $i$ with spin $\sigma\in\left\{\uparrow,\downarrow\right\}$; $\hat{n}_{i\sigma}\equiv \hat{c}^\dag_{i\sigma} \hat{c}_{i\sigma}$ denotes the occupation number operator, $U\geq 0$ is the repulsive on-site interaction, $t$ the nearest-neighbor in-plane hopping, $t^\prime$ the inter-layer hopping, and $\mu$ the chemical potential. By $\left\langle i,j\right\rangle$  we denote nearest neighbors $i$, $j$ within a plane and by $\left\langle i,j\right\rangle^{\prime}$ nearest neighbor pairs in adjacent planes.

We investigate the case $t\geq t^\prime\geq 0$. Both lattices are bipartite and the model is thus particle-hole symmetric with half filling corresponding to $\mu=U/2$. The simply stacked square lattice in the regime of weakly coupled chains, $t^\prime\geq t\geq 0$, was studied in Ref.~\onlinecite{AnisotropicHubbard14}. Note that this simply stacked honeycomb lattice does not correspond to the lattice of graphite, where adjacent ayers are shifted relative to each other.

The non-interacting bandwidth of the studied lattices is $W=2Z_t t+4t^\prime$, where $Z_t$ denotes the in-plane coordination number $Z_t=3$ for the stacked honeycomb lattice, and $Z_t = 4$ for the stacked square lattice.

\section{Method}
\label{sec:Method}

We use the dynamical cluster approximation~\cite{DCA2001} (DCA) method to simulate the Hubbard model on both lattices. The stacked square lattice has a single site per unit cell and may be simulated by the standard DCA method. The simply stacked honeycomb lattice is simulated by a generalization of DCA formulated for an $\ell$-site unit cell, which is explained in detail in Sec. \ref{sec:AppendixMultisiteDCA}.

For simulations of the paramagnetic phase we use the two-site unit cell depicted in the left panel of Fig.~\ref{fig:lattices}. The basis vectors of the simply stacked honeycomb lattice are ${\bf a}_{1}=\left(\frac{3}{2},\frac{\sqrt{3}}{2},0\right)$, ${\bf a}_2=(0,\sqrt{3},0)$, ${\bf a}_3=(0,0,1)$, and the intracell vectors are ${\bf r}_A={\bf 0}$, ${\bf r}_B = (1,0,0)$. The basis vectors of the stacked square lattice are the unit vectors in $x$, $y$, and $z$ direction, ${\bf a}_1=(1,0,0)$, ${\bf a}_2=(0,1,0)$, ${\bf a}_3=(0,0,1)$.

We locate the temperature of the N\'eel phase transition by measuring the divergence of the antiferromagnetic susceptibility. We found this method to be superior to allowing for translational symmetry breaking by doubling the unit cell and measuring the staggered magnetization directly. The reason is a critical slowing down of the DCA self-consistency loop close to the phase transition. Details are presented in  Appendix~\ref{sec:AppendixSusceptibility}.

Most of the clusters utilized in the study respect the three-fold (four-fold) rotational symmetry around the vertical axis of the stacked honeycomb (square) lattice. The aspect ratio of the clusters is chosen to be similar to the anisotropy $t/t^\prime$. Since non-bipartite clusters may cause artificial frustration at low temperature, we used them only for equation of state (EOS) calculations above the N\'eel temperature. In particular, we used simply stacked single and triple layered clusters, which are non-bipartite in the direction of the weak hopping $t^\prime$. Tables listing the clusters are given in Appendix~\ref{sec:AppendixClusters}. 

The impurity solver employed in the study is the continuous time auxiliary field quantum Monte Carlo solver~\cite{CTAUX2008} with sub-matrix updates.\cite{CTAUXsubmatrixUpdates2011} %The sign problem is absent at half filling and simulation in neighborhood of the half-filled case are feasable as well.

% Extrapolation with cluster size 

\subsection{Framework for multisite unit cells}
\label{sec:MultisiteFormulation}
In this subsection we present the notation for description of a general non-Bravais lattice consisting of $\ell$-site unit cells.
The general translationally invariant non-interacting Hamiltonian on such a lattice is of the form
\begin{eqnarray}
	\hat{H}^0 &=& -\sum_{%\mathclap{
	%{\bf r},{\bf \Delta},\alpha,\alpha^\prime,\sigma} \tilde{t}_{{\bf \Delta}\alpha\alpha^\prime} \hat{c}_{{\bf r}\alpha\sigma}^\dagger \hat{c}_{{\bf r}+{\bf \Delta}\alpha^\prime\sigma}  + \hat{H}_\textrm{int} .
	{\bf r},{\bf r}^\prime,\alpha,\alpha^\prime,\sigma} \tilde{t}_{({\bf r}^\prime-{\bf r})\alpha\alpha^\prime} \hat{c}_{{\bf r}\alpha\sigma}^\dagger \hat{c}_{{\bf r}^\prime\alpha^\prime\sigma}  .
\end{eqnarray}
The site position is described by a pair of cell realspace position ${\bf r}$ (${\bf r}^\prime$) and of an sublattice index $\alpha$ ($\alpha^\prime$).
The cell realspace position is an integer linear combination of the lattice basis vectors ${\bf a}_i$, the sublattice index is a number from $\left\{1,\:2,\:\ldots,\:\ell\right\}$, and $\sigma\in\left\{\uparrow,\downarrow\right\}$.

We use a Fourier transformation (FT) to relate creation operators in reciprocal space and real space,
\begin{equation}
	\hat{c}_{{\bf r}\alpha\sigma}^\dagger = \frac{1}{\sqrt{L}} \sum_{{\bf k}} e^{-i {\bf k}\cdot({\bf r}+{\bf r}_\alpha)} \hat{c}_{{\bf k}\alpha\sigma}^\dagger , \label{eq:FT in space}
\end{equation}
where ${\bf r}_\alpha$ is the intracell position vector of site $\alpha$, and $L$ is the number of unit cells.
The non-interacting Green's function in Matsubara representation in reciprocal space is then conveniently given in a matrix form,
\begin{eqnarray}
	\left[G^0_\sigma({\bf k},i\omega_n)\right]_{\alpha\alpha^\prime} &\equiv & -\int_0^\beta\mathrm{d}\tau\:e^{i\omega_n\tau}\left\langle \hat{c}_{{\bf k}\alpha\sigma}(\tau) \hat{c}^\dag_{{\bf k}\alpha^\prime \sigma} \right\rangle , \\
	&=& \left[\left(i\omega_n \mathbb{1}_{\ell}+\tilde{T}_{\bf k}\right)^{-1}\right]_{\alpha\alpha^\prime} ,
\end{eqnarray}
where $\tilde{T}_{\bf k}$ is an $\ell\times\ell$ matrix composed of the Fourier transformed $\tilde{t}_{{\bf \Delta}\alpha\alpha^\prime}$, 
\begin{equation}
	\left(\tilde{T}_{\bf k}\right)_{\alpha\alpha^\prime}=\tilde{t}_{{\bf k}\alpha\alpha^\prime}=e^{i{\bf k}\cdot({\bf r}_{\alpha^\prime}-{\bf r}_{\alpha})} \sum_{\bf \Delta} e^{i {\bf k}\cdot{\bf \Delta}}\: \tilde{t}_{{\bf \Delta}\alpha\alpha^\prime}.
\end{equation}

\subsection{DCA method for multisite unit cells}
\label{sec:AppendixMultisiteDCA}
We employ a generalization of the DCA method applicable for translationally invariant models with unit cells containing $\ell$ sites, introduced in studies of the (extended) Hubbard model on the honeycomb lattice~\cite{LiebschDCAvsCDMFT2013,DCAHoneycomb} and to allow study of symmetry broken phases in a two-dimensional Kondo lattice model. \cite{DissOnDCA2010}

The DCA patch is the Brilluoin zone of the superlattice defined by the cluster. In order to make it compact in terms of dispersion variations, we choose it to be the Wigner--Seitz cell constructed with distance function $\Vert {\bf k}\Vert ^2=(k_x^2+k_y^2)+({t^\prime}/t)^2k_z^2$ taking into account the anisotropy.

DCA approximates the lattice self energy by the patch-wise constant cluster self energy $\Sigma$. The mapping of the lattice onto a cluster is in the reciprocal space expressed as patch averaging,
\begin{equation}
	G({\bf K}) = \frac{1}{\Omega}\int_{\textrm{patch}} \mathrm{d}\tilde{\bf k}\: \left[G^0({\bf K}+\tilde{{\bf k}})^{-1} - \Sigma({\bf K}) \right]^{-1} , \label{eq:DCAmapping}
\end{equation}
where $G({\bf K})$ denotes the interacting cluster Green's function, which is to be determined selfconsistently along with the non-interacting cluster Green's function $\mathscr{G}^0({\bf K})$. We suppressed for shortness the Matsubara frequencies $i\omega_n$ and spin projection $\sigma$ in all quantities. ${\bf K}$ denotes a cluster reciprocal vector, and $\Omega$ the patch volume. Notice that the mapping condition is formally identical to the DCA mapping for Bravais lattices, with corresponding quantities being $\ell\times\ell$ matrices instead of scalars.

%DCA for $\ell$-site cells onto the cluster has the same form as the DCA for Bravais lattices, with corresponding quantities being $\ell\times\ell$ matrices instead of scalars. {\color{red} \bf [MT: This cannot be understood in its present form and needs to be expanded]}. 
The mapping in Eq.(\ref{eq:DCAmapping}) is solved iteratively by the self consistency procedure summarized by the steps:
\begin{eqnarray}
	\Sigma({\bf K}) &=&  \mathscr{G}^0({\bf K}) ^{-1} - G({\bf K}) ^{-1} , \\	
	\bar{G}^{\textrm{lat}}({\bf K}) &=& \frac{1}{\Omega}\int_{\textrm{patch}} \mathrm{d}\tilde{\bf k}\: \left[G^0({\bf K}+\tilde{{\bf k}})^{-1} - \Sigma({\bf K}) \right]^{-1} , \label{eq:DCAavereging} \\
	\mathscr{G}^0({\bf K}) &=&  \left[\bar{G}^{\textrm{lat}}({\bf K})^{-1} + \Sigma({\bf K})\right]^{-1},
\end{eqnarray}
%where we suppressed the Matsubara frequencies $i\omega_n$ and spin projection $\sigma$ in all quantities; $\mathscr{G}^0({\bf K})$ ($G({\bf K})$) denotes the non-interacting (interacting) cluster Green's function, 
where $\bar{G}^{\textrm{lat}}({\bf K})$ denotes the patch averaged interacting lattice Green's function.
% $\Sigma({\bf K})$ for the self energy, and $\Omega$ denotes the patch volume.

Note that the averaging step of Eq.~(\ref{eq:DCAavereging}) makes the DCA with multisite cells sensitive to the specific choice of FT. 
Omitting the phase factors $e^{-i{\bf k}\cdot{\bf r}_\alpha}$ in Eq. (\ref{eq:FT in space}) in the FT, as done in Ref.~\onlinecite{DissOnDCA2010}, leads, in general, to different results for any finite size cluster. Our choice of the form of FT in Eq. (\ref{eq:FT in space}) produces a slower varying $G^0({\bf k})$ within a patch, thus effectively reducing the mean-field effects associated with the averaging. The difference originates from the slower varying off-diagonal elements of $\tilde{T}_{\bf k}$.
Another advantage of the FT form shows up if the problem possesses point group symmetries, as these then directly propagate to the Green's functions.

\section{Results}
\label{sec:Results}

We compute the EOS and further properties -- the energy, entropy, density, nearest-neighbor spin correlation, and the double occupancy -- of the model of Eq.~(\ref{eq:Ham}) by using DCA and extrapolating to the thermodynamic limit according to $L^{-2/3}$ (see \onlinecite{Fuchs:2011ch,Jarrell:2002} for details).
We restrict the calculations to fillings $n \leq 1$ per site, as the results for $1<n\leq 2$ are related to those for $n \leq 1$ via particle-hole symmetry.

\subsection{Free energy and entropy}

The entropy per site $s$ is estimated by numeric integration at fixed $U$ and $\mu$, starting from a high temperature $T_u/t=50$,
\begin{eqnarray}
	s(T) &=& s(T_u) + \frac{f(T)}{T} - \frac{f(T_u)}{T_u} + \int^{T_u}_{T} \frac{f(T^\prime)\:\mathrm{d}T^\prime}{{T^\prime}^2} . \label{eq:entropy_integration}
\end{eqnarray}
Here, $f(T) = e(T)-\mu n(T)$, with $e$ the energy per site and $n$ the density per site. The value of $s(T_u)$ is obtained from a high-temperature series expansion,
\begin{eqnarray}
	s(T_u) &=& \ln 4 - \frac{1}{2 T_u^2}\left[\frac{U^2}{16}+\frac{(\mu-U/2)^2}{2}+\frac{Z_t t^2+2t^{\prime 2}}{2} \right]  \nonumber\\
	&& + \frac{U \left(\mu-U/2\right)^2}{8T_u^3}   + O(T_u^{-4}) .
\end{eqnarray}
The expression for $e(T)$ is explicitly given in Eq.~(\ref{eq:energy_per_site}).
We provide the EOS at half filling for interactions $U/t=1,2,3,4,6,8$ and anisotropies $t/t^\prime=1,2,4,8$ for both stacked honeycomb and stacked square lattice. For the stacked honeycomb lattice we choose two parameter sets $(U/t,t/t^\prime)=(6,6)$ and $(4,4)$, at which we obtain the EOS at a wide range of fillings. For the stacked square lattice we do the same at $(U/t,t^\prime/t)=(6,6)$. The EOS, being one of the main outcomes of the study, is presented in the Supplemental Material.\cite{supplementary}

\begin{figure}%[t]
\centering
\includegraphics[width=8cm]{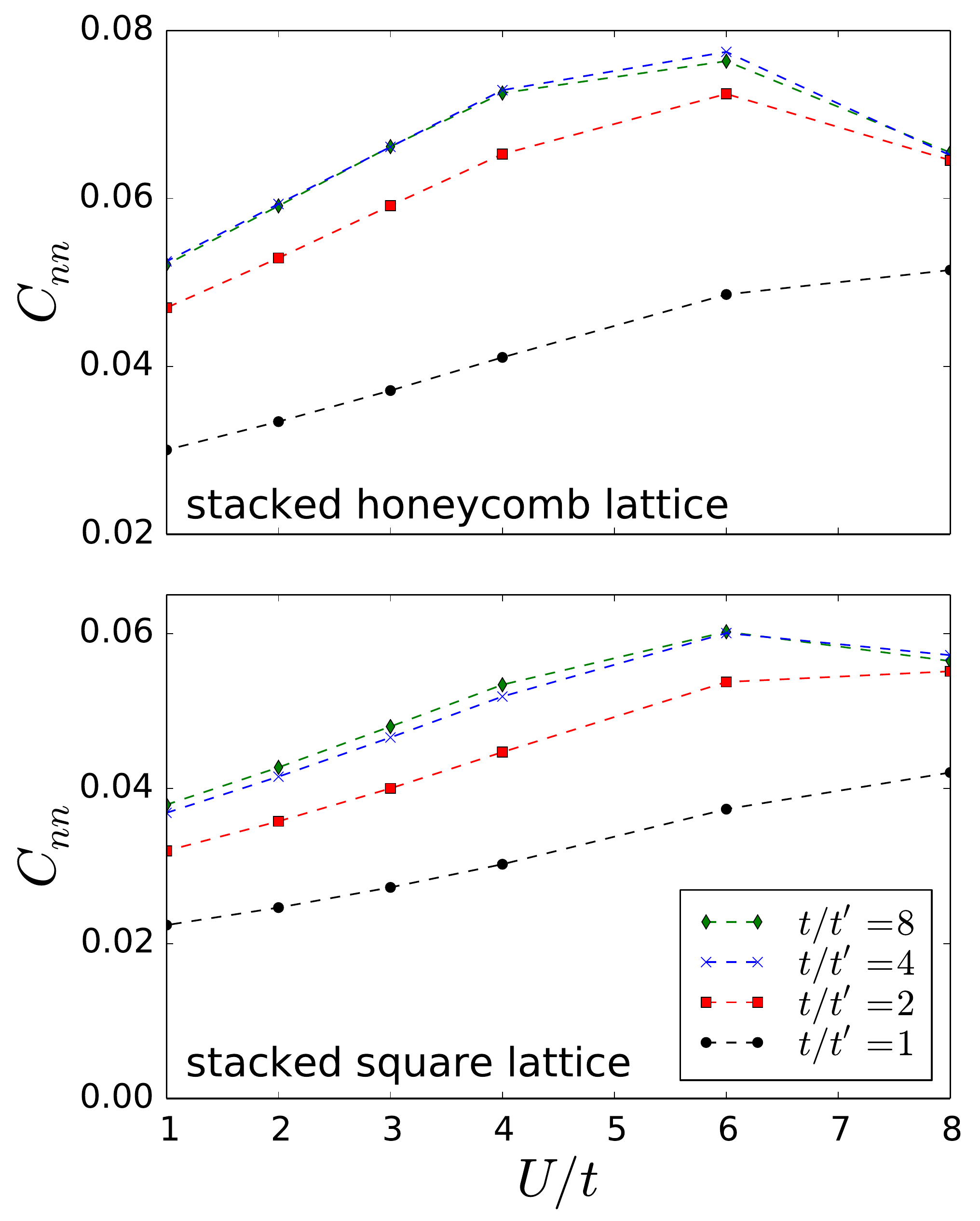}
\caption{The nearest neighbor in-plane spin correlation $C_{nn}$ for the stacked honeycomb lattice (top) and stacked square lattice (bottom) as a function of the interaction strength $U/t$ for various anisotropies $t^\prime/t$ at half filling ($n=1$, $\mu=U/2$) and at an entropy $s=0.7$. $C_{nn}$ shows an enhancement in the anisotropic case $t/t^\prime > 1$.}
\label{fig:spincorr_vs_U_anisotropy}
\end{figure}

\subsection{Spin correlations}

We calculated the nearest-neighbor spin correlations,\cite{ShortRangeCorrelationsExp} which capture the onset of magnetic ordering and have proven to be a suitable observable for calculated the temperature of the system.\cite{Fuchs:2011ch,AnisotropicHubbard14} We specifically calculate the equal-time in-plane nearest-neighbor spin correlations
\begin{eqnarray}
	C_{nn} &=& -\frac{2}{Z_t L \ell}\sum_{\left\langle i,j \right\rangle}\left\langle \hat{S}^z_{i} \hat{S}^z_{j}\right\rangle ,
\end{eqnarray}
where the sum runs over in-plane nearest-neighbor pairs (coupled by the strong hopping $t$), $\hat{S}^z_{i} = \frac{1}{2}(\hat{n}_{i\uparrow} -  \hat{n}_{i\downarrow})$, $L$ is the number of cells and the average $\left\langle \hat{S}^z_{i} \hat{S}^z_{j}\right\rangle$ is measured directly on the cluster. We show the spin correlations for the stacked honeycomb and stacked square lattice in~Fig.~\ref{fig:spincorr_vs_U_anisotropy}. The  data shown was calculated for a homogeneous system at half filling and at a fixed entropy per particle. $C_{nn}$ shows similar behavior with interaction strength and anisotropy for both lattices, with an approximative amplification by a factor $4/3$ in the stacked honeycomb lattice. The factor $4/3$ is the ratio of strong hopping coordination numbers $Z_t$ for these lattices. The maxima are at similar interaction strengths if interactions are measured in units of the bandwidth $W$. Qualitatively, the observed behavior is captured by the second order high-temperature expansion, which is (at half filling) given by
\begin{equation}
	C_{nn}^{(2)}(s) = \frac{2(\ln 4-s)t^2}{8(Z_t t^2 +2{t^\prime}^2)+U^2}. % + O\left((\ln 4 -s)^2\right) .
\end{equation}
Quantitatively, the second order high-temperature estimate of $C_{nn}(T)$ is reliable only for $T/t\gtrapprox 3$, corresponding to an entropy per site $s$ well above $1$. Note that Fig.~\ref{fig:spincorr_vs_U_anisotropy} is calculated for $s=0.7$, which is close the lowest experimentally realizable value at half filling.\cite{AnisotropicHubbard14} Noticeably, the sum of $C_{nn}^{(2)}(s)$ over all bonds $\left\langle i,j\right\rangle$ is independent of the lattice properties if $U$ is scaled according to the root of the second moment of the non-interacting density of states $D_2=Z_t t^2+2{t^\prime}^2$.
%\begin{equation}
%	\frac{-2}{N}\sum_{\left\langle i,j\right\rangle} \left\langle \hat{S}^z_{i} \hat{S}^z_{j}\right\rangle = \frac{\ln 4 -s}{8+U^2/D_2}.
%\end{equation}

$C_{nn}$, as an experimentally measurable quantity,~\cite{ShortRangeCorrelationsExp} may serve as a sensitive thermometer in the temperature range $T/t\sim 1$ if compared with the EOS we provide. The enhancement due to anisotropy raises the signal and renders the measurement more precise.

\subsection{Trap effects}

\begin{figure}
\centering
\includegraphics[width=8.5cm]{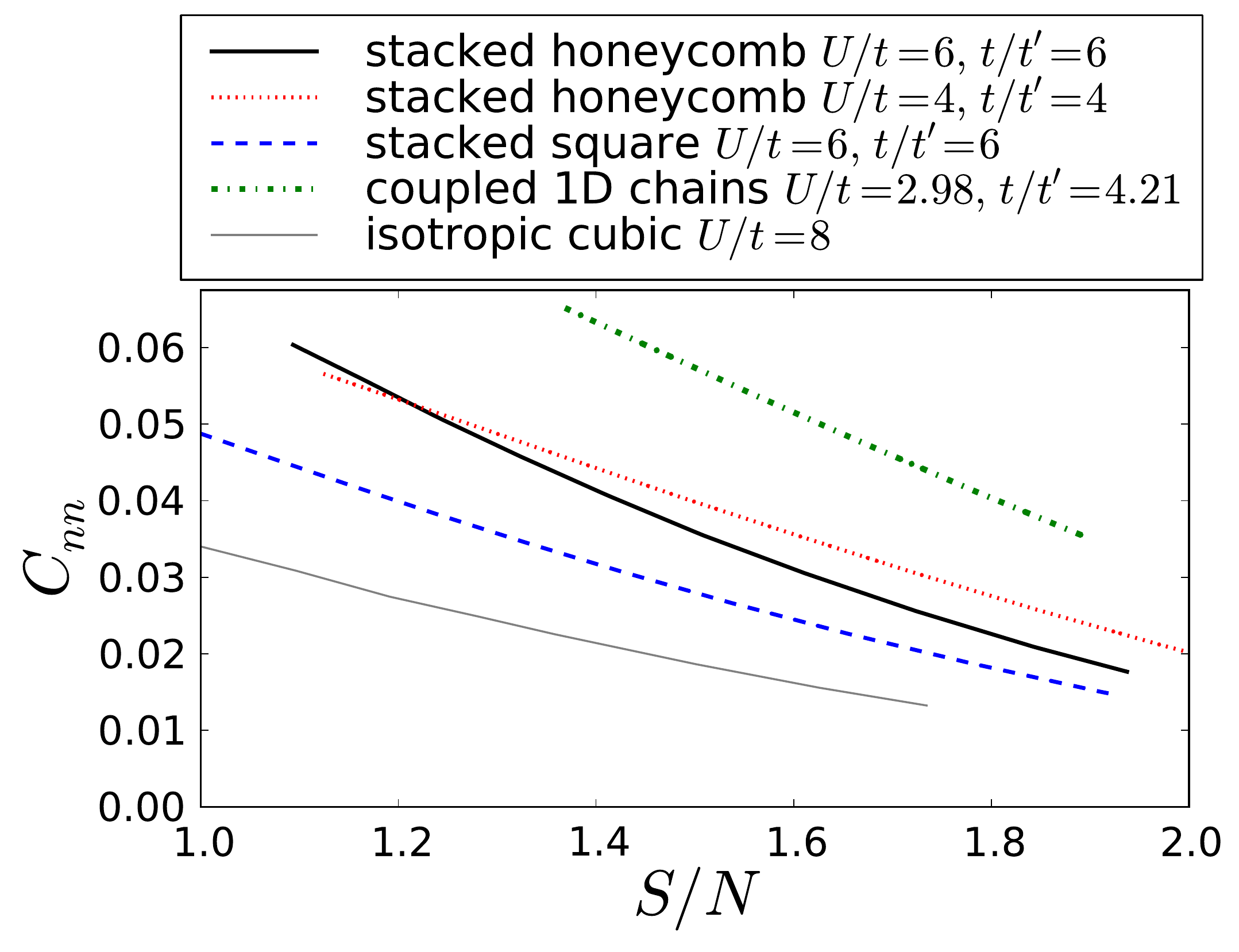}
\caption{Average spin correlation $C_{nn}$ per particle plotted as a function of the entropy per particle $S/N$ in a quadratic trap with chemical potential adjusted to obtain half filling ($n=1$) in the trap center. The data for the coupled 1D chains and isotropic cubic lattice is taken from Refs.~\onlinecite{AnisotropicHubbard14,Fuchs:2011ch}.}
\label{fig:LDA}
\end{figure}

\begin{figure}%[t]
\centering
\includegraphics[width=8cm]{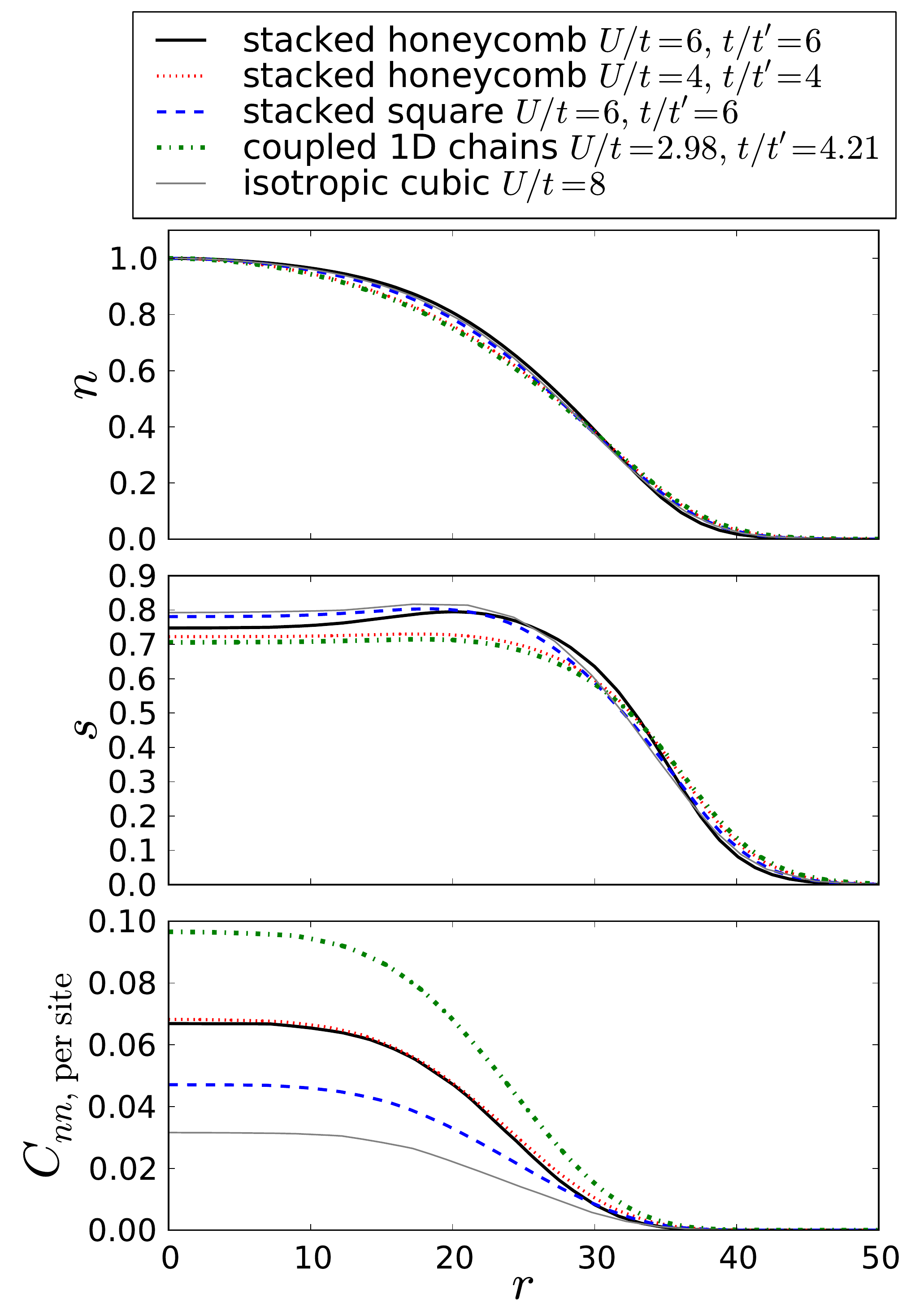}
\caption{Density $n$, entropy $s$, and $C_{nn}$ profile in a quadratic trap with the particle number fixed to $N=10^5$ and half filling in its center. The radius $r$ is given in units of lattice sites. The entropy per particle $S/N$ is set to $1.4$, which is an experimentally achievable value~\cite{AnisotropicHubbard14}.
%The temperatures are $T=0.6966, 0.7639, 0.8471, 0.5799$ in units of strong hopping for stacked hexagonal lattice at $(U/t,t/t^\prime)=(4,4),(6,6)$, stacked square lattice at $U=6t,t/t^\prime=6$ and coupled 1D chains at $U=2.98t,t/t^\prime=4.21$.
The data for the coupled 1D chains and isotropic cubic lattice is taken from Ref.~\onlinecite{AnisotropicHubbard14,Fuchs:2011ch}.}
\label{fig:LDAatS1.4}
\end{figure}

In experiments ultracold atoms are  confined by a trapping potential, which may be modeled by a local density approximation (LDA)
at currently experimentally accessible temperatures.\cite{Scarola:2009cb,Zhou:2011dj} 
As  confining potential we take a quadratic function $V({\bf x})$ with minimum in the trap center. The chemical potential $\mu$ we choose such that the system is half-filled in the trap center. Assuming a large lattice we use a continuous approximation instead of discrete summation over lattice sites and obtain a trap-averaged quantity  $Q$ as
\begin{equation}
	Q = \int \mathrm{d}^3 {\bf x}\  q\left(\mu-V({\bf x});T\right),
\end{equation}
where $q$ is the density of the quantity of interest in a homogeneous system. With this definition, the quantity $Q$ per particle, $Q/N$, is independent on the specific parameters of the quadratic potential. For the LDA calculations we need the EOS at low filling, which we approximate by the EOS of the corresponding non-interacting system. Fig.~\ref{fig:LDA} shows the trap-averaged $C_{nn}$ and Fig.~\ref{fig:LDAatS1.4} shows profiles of density, entropy per site, and $C_{nn}$ for the stacked honeycomb and square lattice and the $1$D coupled chains. For Fig.~\ref{fig:LDAatS1.4} we assume isotropic $V({\bf x})$. The distance from the trap center we denote by $r$. In the studied temperature regime, coupled 1D chains~\cite{AnisotropicHubbard14} show the largest spin correlation.\footnote{For the 1D chain the $C_{nn}$ is the nearest-neighbor spin correlation in direction of the strong hopping, see definition in Ref.~\onlinecite{AnisotropicHubbard14}.} The density and entropy distributions differ only marginally. In the lower panel of Fig.~\ref{fig:LDAatS1.4} we observe that $C_{nn}$ at half filling ($r=0$) is roughly proportional to the inverse strong hopping coordination number $Z_t^{-1}$. This effect might be qualitatively explained by the different energy scales of the hoppings -- the simulations are performed at a high temperature relative to the weak hopping $t'$, but the temperature is comparable with the strong hopping $t$. Thus the antiferromagnetic short-range correlations tend to build up in the strong hopping directions (in-plane) and the singlet formation is facilitated by lower $Z_t$.

\subsection{Double occupancy and adiabatic cooling}

Of further experimental interest are ways to cool the particles, to provide access to interesting low temperature phenomena. As proposed in Ref.~\onlinecite{AdiabaticCooling05},  fermions in an optical lattice may be adiabatically cooled by increasing of the interaction strength if the double occupancy $D=\frac{1}{N} \langle n_{i\uparrow} n_{i\downarrow} \rangle$ shows an increase upon cooling at fixed density $n$. 
%The importance of the procedure was further discussed in~\cite{AdiabaticCooling07} with a rather pessimistic conclusion. 
The interaction driven adiabatic cooling was experimentally utilized for a $SU(6)$ Hubbard model.~\cite{PomeranchukCooling12} For the Hubbard model  in the context of optical lattice experiments, the presence of the effect was numerically observed both for square and honeycomb lattices.\cite{Thermodynamics2D_11,ThermodynamicsOfHoneycomb2012}

\begin{figure}
\centering
\includegraphics[width=8cm]{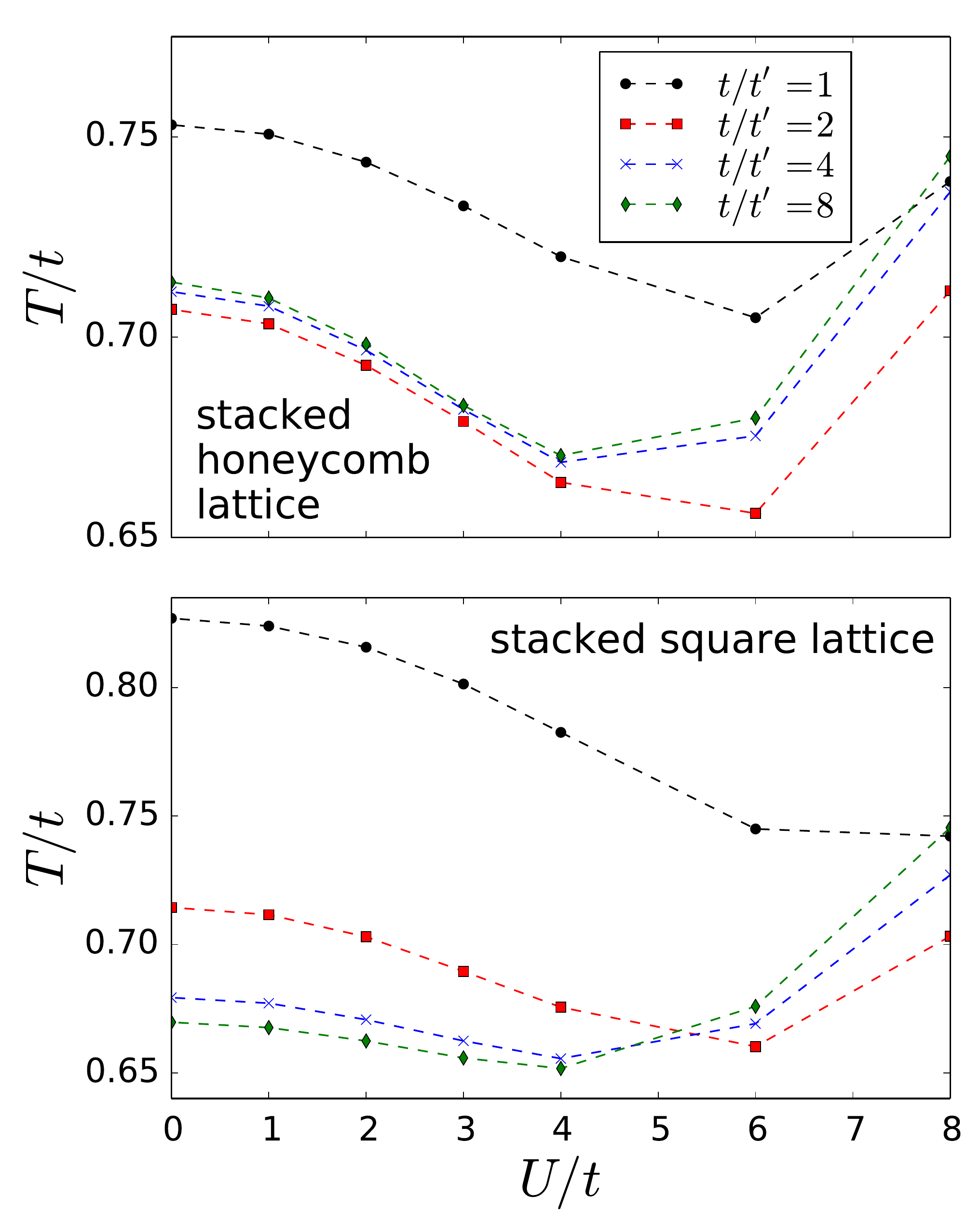}
\caption{Temperature $T/t$ plotted as a function of $U/t$ for the stacked honeycomb (top) and stacked square lattice (bottom) at half filling and at entropy per site $s=0.7$ for various anisotropies $t/t^\prime$. An adiabatic increase of $U/t$ from $0$ up to a parameter-specific $U$ induces cooling in all cases.}
\label{fig:adiabatic_cooling}
\end{figure}

\begin{figure}%[t]
\centering
\includegraphics[width=8.5cm]{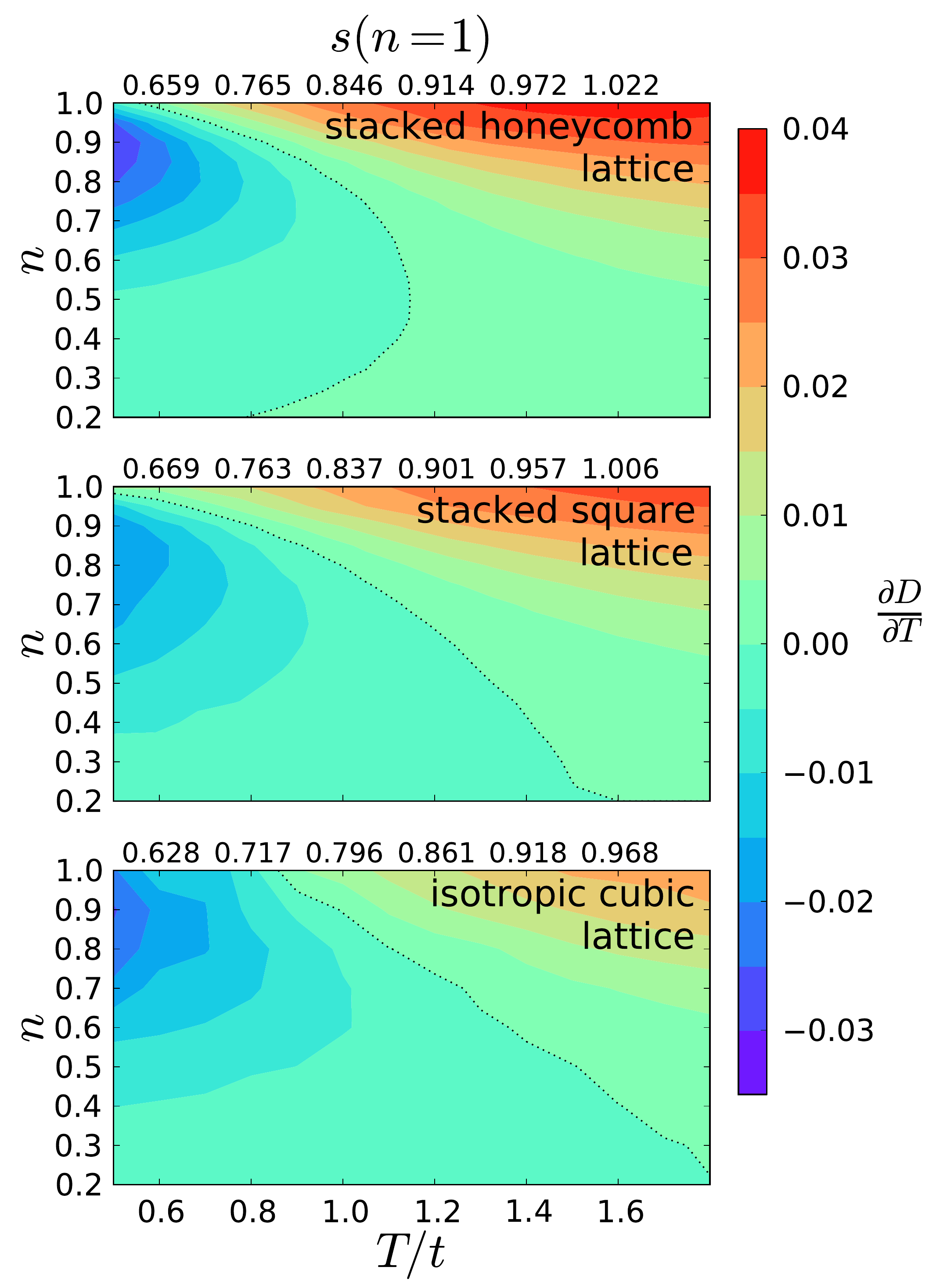}
\caption{Negative values of $\frac{\partial D}{\partial T}$ (taken at fixed density $n$) indicate presence of the effect of adiabatic cooling upon interaction increase. The corresponding quantity is plotted as a function of $T$ and $n$ for the stacked honeycomb (top) and stacked square lattice (middle) at $U/t=6$ and $t/t^\prime=6$. For comparison we show the same quantity for isotropic cubic lattice at $U/t=6$ in the bottom panel, using data from Ref.~\onlinecite{Fuchs:2011ch}. The regions of positive and negative $\frac{\partial D}{\partial T}$ are separated by dotted line. For a fair comparison we add upper axis with entropy per site of half filled system, $s(n=1)$, at temperature given by the temperature axis common to all plots.}
\label{fig:double_occupation}
\end{figure}

We here investigate this effect for the stacked lattices in a homogeneous system.
Figure \ref{fig:adiabatic_cooling} shows the adiabatic cooling effect at half filling and at entropy per site $s=0.7$ at a range of anisotropies, with cooling persisting up to $U/t\approx 6$. Alternatively it is possible to start from large interactions and decrease $U$; however, in that case $T/U$ may increase. 
The cooling is present only at sufficiently low entropies, $s \lessapprox 0.8$, and it is accompanied by an approximate maximization of $C_{nn}$ according to the Fig.~\ref{fig:spincorr_vs_U_anisotropy}(top). 
%It is noticeable that the interaction increase to the optimal $U$ minimizing $T/t$ is accompanied by approximate maximization of $C_{nn}$ according to the Fig.~\ref{fig:spincorr_vs_U_anisotropy}(a). 
The stacked square lattice shows the largest effect in the isotropic limit, equivalent to the cubic lattice.

Figure \ref{fig:double_occupation} shows the slope of $D(T)$ away from half filling. Cooling here appears at even higher temperature than in the half filled case.
This might be utilized to transfer entropy from the region with half filling to less densely occupied regions in the trap.
While realistic cooling design was discussed in Refs.~\onlinecite{CoolingByShaping,CoolingByInteraction}, we only note that the low density regions show large entropy per particle and thus they may store a large portion of the total entropy. Fig.~\ref{fig:double_occupation} shows that there are no qualitative but only subtle quantitative differences in between the examined lattices with respect to the presence and strength of the cooling phenomenon. The adiabatic cooling effect $\left( \frac{\partial T}{\partial U} \right)_{s}$  is proportional to the inverse of the specific heat and to $\left( \frac{\partial D}{\partial T} \right)_{U}$. As the magnitude of the latter does not show great differences among the investigated lattices, the cooling effect is of comparable strength with some enhancement in the case of stacked honeycomb lattice at density near to the half filling.

\subsection{N\'eel transition}
The entropy per particle at the N\'eel temperature $T_N$ is expected to decrease for large anisotropies, in accordance with the Mermin--Wagner--Hohenberg theorem.\cite{MerminWagnerTheorem1966,Hohenberg1967} We investigate the N\'eel transition for half filling only. In order to identify the lattice with the largest entropy per site at the N\'eel transition, $s_N$, we therefore focused on smaller anisotropies in this part. Since any mean-field theory overestimates ordering, the $T_N$ and $s_N$ for a specific cluster provides an upper bound of the corresponding quantities in the thermodynamic limit. For an unbiased estimate we localize the transition temperature for several clusters and extrapolate the transition temperature $T_N$ as suggested in Ref.~\onlinecite{Jarrell:2005ec}, using the critical exponent $\nu=0.71$ for the 3D Heisenberg model.\cite{Sandvik1998} $T_N$ and $s_N$ calculated for the different systems studied in this paper are summarized in the Tab.~\ref{tab:T_Neel}.

\begin{table}
\centering
\begin{tabular}{| c | c | c | c | c  c }
	\hline
	\multicolumn{2}{|c|}{} & \multicolumn{2}{|c|}{stacked square} & \multicolumn{2}{|c|}{stacked honeycomb} \\
	\hline
	$t/t^\prime$ & $U/t$ & $T_N/t$ & $s_N$ & $T_N/t$ & \multicolumn{1}{|c|}{$s_N$} \\
	\hline
	\multirow{3}{*}{1} & 4 &  0.1955(25) & 0.223(18) & 0.206(1) & \multicolumn{1}{|c|}{}  \\
	& 6 & 0.324(2) & 0.430(8) & 0.292(4) & \multicolumn{1}{|c|}{0.405(35)} \\
	& 8 & 0.3595(83) & 0.487(23) & 0.299(2) & \multicolumn{1}{|c|}{0.33(10)} \\
	\hline
	\multirow{3}{*}{2} & 4 & 0.206(2) & 0.313(49) & 0.173(4) & \multicolumn{1}{|c|}{}  \\
	& 6 & 0.293(4) & 0.438(39) & 0.239(11) & \multicolumn{1}{|c|}{0.28(8)} \\
	& 8 & 0.294(8) & 0.41(6) & 0.205(21) & \multicolumn{1}{|c|}{} \\
	\hline
	\multirow{3}{*}{4} & 4 & 0.202(3) & 0.30(7) &  & \\
	& 6 & 0.241(5) & 0.27(11) & & \\
	& 8 & 0.219(8) & & & \\
	\cline{1-4}
	\multirow{2}{*}{8} & 4 & 0.185(2) & & & \\
	& 6 & 0.208(7) & & & \\
	\cline{1-4}
\end{tabular}
\caption{N\'eel temperatures and entropies for both examined stacked lattices. For stacked square lattice we studied wide range of anisotropies as those may be of interest with respect to undoped high-$T_c$ superconductor parent materials.
Missing $s_N$ entries indicate that we did not integrate the entropy down to $T_N$.
The isotropic cubic lattice data for $U/t=4,8$ is from Ref.~\onlinecite{AnisotropicHubbard14}.
}
\label{tab:T_Neel}
\end{table}

\section{Conclusion}
\label{sec:Conclusion}
For both lattice structures we calculated the EOS at half filling for anisotropies $1\leq t/t^\prime\leq 8$ and interactions $1\leq U/t \leq 8$, and for a few parameter sets we obtained the EOS for a wide range of fillings. The EOS data is provided in the Supplemental Material.\cite{supplementary} It allows for LDA based trap-specific calculations, suitable for calibration and optimization of ultracold atoms experiments. In particular it may be used for estimating the amount of heating during the lattice loading and for estimating the temperature and entropy, if the experiment provides a temperature sensitive measurement and the value of the initial entropy.\cite{AnisotropicHubbard14}

For  stacked lattices we found enhanced short-range in-plane correlations for experimentally accessible temperatures of the order $t$. We investigated the possibility of interaction-driven adiabatic cooling, which is present in the studied systems and may contribute to the progress in cooling ultracold atoms. We computed the N\'eel temperature for both investigated stacked lattices. Among them, the lattice with highest critical entropy per particle at half filling is the conventional isotropic cubic lattice with interaction $U/t\approx 8$. For stacked square lattice we investigated the N\'eel temperature also for larger anisotropies, up to $t/t^\prime=8$, as the stacked square lattice at half filling is an effective model for the undoped high-$T_c$ superconductors.

\begin{acknowledgments}
We thank Lei Wang, Daniel Greif, and Jan Gukelberger for useful discussions, and Michael Messer and Gregor Jotzu for careful reading of the manuscript. This work has been supported by the European research Council through ERC Advanced Grant SIMCOFE and by the Swiss National Science Foundation through the National Competence Center for Research QSIT. The calculations were based on the ALPS libraries~\cite{ALPS1.3:2007,ALPS2.0:2011} and were performed on the Brutus and Monch clusters of ETH Zurich and at the Texas Advanced Computing Center (TG-DMR130036).
\end{acknowledgments}

\appendix

\section{Susceptibility measurement}
\label{sec:AppendixSusceptibility}
This section provides details on the susceptibility measurement in the DCA with $\ell$-site cell. It is a generalization of the susceptibility measurement described in Ref.~\onlinecite{QClusterTheoriesReview2005}. We stick to the naming and denotations used therein.

Our interest is in the static (zero bosonic frequency $\nu$) susceptibility corresponding to the staggered magnetization operator,
\begin{equation}
	\hat{m}=\frac{1}{L \ell}\sum_{{\bf r},\alpha,\sigma} e^{i{\bf Q}\cdot{\bf r}}\:f_\alpha\:\sigma\:  \hat{n}_{{\bf r}\alpha\sigma} , \label{eq:MultisiteMagnetizationOperator}
\end{equation}
with $\sigma=\pm 1$.
For the stacked square lattice we use a description with a single site per unit cell and ${\bf Q}$ is then the antiferromagnetic reciprocal vector $\left(\pi,\pi,\pi\right)$, and $f_{A}=1$.
For the stacked honeycomb lattice we performed simulations in the paramagnetic regime with $2$-site unit cell depicted in Fig.~\ref{fig:lattices}(left) and we measured the susceptibility at ${\bf Q}=\left(0,0,\pi\right)$; with sublattice factors $f_A=1$, $f_B=-1$.

In the following we assume ${\bf Q}\neq {\bf 0}$ to simplify the formulas, as both cases satisfies that condition. We assume the ${\bf Q}$ vector to be contained in the set of cluster reciprocal vectors.

We follow these steps:
\begin{enumerate}
	\item The two-particle cluster Green's function $\chi_c({\bf Q},i\nu)$ is measured using the CTAUX impurity solver~\cite{CTAUX2008} and Wick's theorem.
	\item The irreducible cluster vertex $\Gamma_c({\bf Q},i\nu)$ is then obtained via Bethe--Salpeter equation,
		\begin{equation}
			\Gamma_c({\bf Q},i\nu)=\chi^0_c({\bf Q},i\nu)^{-1}-\chi_c({\bf Q},i\nu)^{-1}.
		\end{equation}
	\item We approximate the irreducible lattice vertex $\Gamma({\bf Q},i\nu)$ by the cluster vertex $\Gamma_c({\bf Q},i\nu)$. $\Gamma({\bf Q},i\nu)$ will thus be patch-wise constant in reciprocal space.
	\item The lattice two-particle cluster Green's function $\chi({\bf Q},i\nu)$ is obtained via Bethe--Salpeter equation.
		\begin{equation}
			\chi({\bf Q},i\nu)^{-1}=\chi^0({\bf Q},i\nu)^{-1}-\Gamma({\bf Q},i\nu).
		\end{equation}
		Since the vertex $\Gamma$ is patch-wise constant, we may perform patch averaging to obtain 
		\begin{equation}
			\bar{\chi}({\bf Q},i\nu)^{-1}=\bar{\chi}^0({\bf Q},i\nu)^{-1}-\Gamma({\bf Q},i\nu).
		\end{equation}
	\item The lattice susceptibility is computed from $\bar{\chi}({\bf Q},i\nu)$.
\end{enumerate}
The precise definitions of all quantities are given below.

\begin{widetext}
For the matrices $\chi_{(c)}({\bf Q},i\nu)$, $\Gamma_{(c)}({\bf Q},i\nu)$, and $\chi^0_{(c)}({\bf Q},i\nu)$ we use multiindex notation $K\equiv ({\bf K}\alpha\gamma n\sigma)$. The $KK^\prime$ element of the two-particle cluster Green's function $\chi_{c}({\bf Q},0)$ is defined by
\begin{eqnarray}
	 && \frac{1}{\beta} \int_0^{\beta}\int_0^{\beta}\int_0^{\beta}\int_0^{\beta}\mathrm{d}\tau_1\:\mathrm{d}\tau_2\:\mathrm{d}\tau_3\:\mathrm{d}\tau_4 %\nonumber \\
%	&& \times 
	e^{-i(\omega_n\tau_1-\omega_n\tau_2+\omega_{n^\prime}\tau_3-\omega_{n^\prime}\tau_4)} %\nonumber \\
	%&& \times 
	\left\langle T_\tau c^\dag_{{\bf K}+{\bf Q}\alpha\sigma}(\tau_1)c_{{\bf K}\gamma\sigma}(\tau_2)c^\dag_{{\bf K}^\prime \alpha^\prime \sigma^\prime}(\tau_3)c_{{\bf K}^\prime+{\bf Q}\gamma^\prime \sigma^\prime}(\tau_4) \right\rangle . \qquad%\nonumber \\
	%&&
	 \label{eq:two-particle Green's function}
\end{eqnarray}

The noninteracting cluster susceptibility $\chi^0_c$ and its patch-averaged lattice counterpart $\bar{\chi}^0$ are defined by
\begin{eqnarray}
	\left[\chi_c^0({\bf Q},0)\right]_{KK^\prime}&=& -\beta \delta_{\sigma\sigma^\prime}\delta_{nn^\prime}\delta_{{\bf K}{\bf K}^\prime}  
	 \left[G({\bf K},i\omega_n)\right]_{\alpha\gamma^\prime} \left[G({\bf K}+{\bf Q},i\omega_n)\right]_{\alpha^\prime \gamma} , \\
	\left[\bar{\chi}^0({\bf Q},0)\right]_{KK^\prime}&=& -\frac{\beta \delta_{\sigma\sigma^\prime}\delta_{nn^\prime}\delta_{{\bf K}{\bf K}^\prime}}{\Omega} \int_{\textrm{patch}} \mathrm{d}\tilde{\bf k}  
	 \left[G^{\textrm{lat}}({\bf K}+\tilde{\bf k},i\omega_n)\right]_{\alpha\gamma^\prime} \left[ G^{\textrm{lat}}({\bf K}+{\bf Q}+\tilde{\bf k},i\omega_n)\right]_{\alpha^\prime\gamma} ,  
\end{eqnarray}
with cluster (lattice) Green's function $G$ ($G^{\textrm{lat}}$), and Kronecker delta $\delta_{ij}$.

The static staggered spin susceptibility of the lattice may be obtained by
\begin{eqnarray}
	\chi_{AF} &=& \frac{1}{L\ell\beta^2}\sum_{\sigma,\sigma^\prime} \sigma\sigma^\prime \sum_{\alpha,\alpha^\prime} f_{\alpha} f_{\alpha^\prime}  \sum_{{\bf K},{\bf K}^\prime} e^{i{\bf G}\cdot{\bf r}_\alpha} e^{i{\bf G}^\prime\cdot{\bf r}_{\alpha^\prime}}  
	\sum_{n,n^\prime} \bar{\chi}_{{\bf K}\alpha\alpha n \sigma {\bf K}^\prime \alpha^\prime\alpha^\prime n^\prime\sigma^\prime}({\bf Q},i\nu=0), \label{eq:SusceptibilitySum}
\end{eqnarray}
with patch-averaged two-particle lattice Green's function $\bar{\chi}$ at the reciprocal vector ${\bf Q}$ and at zero frequency $\nu$; and with intracell phase factors,
\begin{eqnarray}
	e^{i{\bf G}\cdot{\bf r}_\alpha} \equiv  e^{-i K_\textrm{repr}({\bf K}+{\bf Q})\cdot{\bf r}_\alpha}  e^{i K_\textrm{repr}({\bf K})\cdot{\bf r}_\alpha}, \qquad
	e^{i{\bf G}^\prime\cdot{\bf r}_{\alpha^\prime}}  \equiv  e^{-i K_\textrm{repr}({\bf K}^\prime)\cdot{\bf r}_{\alpha^\prime}} e^{i K_\textrm{repr}({\bf K}^\prime+{\bf Q})\cdot{\bf r}_{\alpha^\prime}},
\end{eqnarray}
where $K_\textrm{repr}({\bf K})$ is the representative of the cluster reciprocal vector ${\bf K}$ used in the simulation, which may differ from ${\bf K}$ by a reciprocal lattice vector ${\bf G}$.\footnote{The Green's function shifted by a reciprocal vector ${\bf G}$ is given by $G({\bf k}+{\bf G})=U_{\bf G}^{+} G({\bf k}) U_{\bf G}$ with diagonal unitary matrix $\left(U_{\bf G}\right)_{\alpha\alpha^\prime}=\delta_{\alpha\alpha^\prime} e^{i{\bf G}\cdot{\bf r}_\alpha}$.}
\end{widetext}

The frequencies $\omega_{n}$, $\omega_{n^\prime}$ need for any practical use be cut-off at some $\omega_c$. The frequency cut-off is cured by fitting the $\frac{1}{\omega_n^2}$ tail and adding its contribution to the result.\cite{HighfreqTailOfVertex}
The cut-off was validated by comparison of the extrapolated impurity susceptibility, obtained by Eq.~(\ref{eq:SusceptibilitySum}) with $\bar{\chi}$ replaced by $\chi_c$, with the directly measured cluster susceptibility,
\begin{equation}
	\chi_{AF,c} = L\ell \int_0^\beta\mathrm{d}\tau\: \langle\hat{m}(\tau)\hat{m}\rangle, \label{eq:AFsusc}
\end{equation}
with $\hat{m}$ given in Eq.~(\ref{eq:MultisiteMagnetizationOperator}).
Typically we used $\omega_c\approx 5U$.

The DCA, as a mean field based method, displays close to the second order phase transition the mean field critical exponents. We utilize that for a precise location of the transition for each cluster by searching for the intersection of $\chi_{AF}^{-\gamma_{\textrm{mf}}}$ with zero, with the mean field critical exponent $\gamma_{\textrm{mf}}=1$.

\begin{figure}%[t]
\centering
\includegraphics[width=8cm]{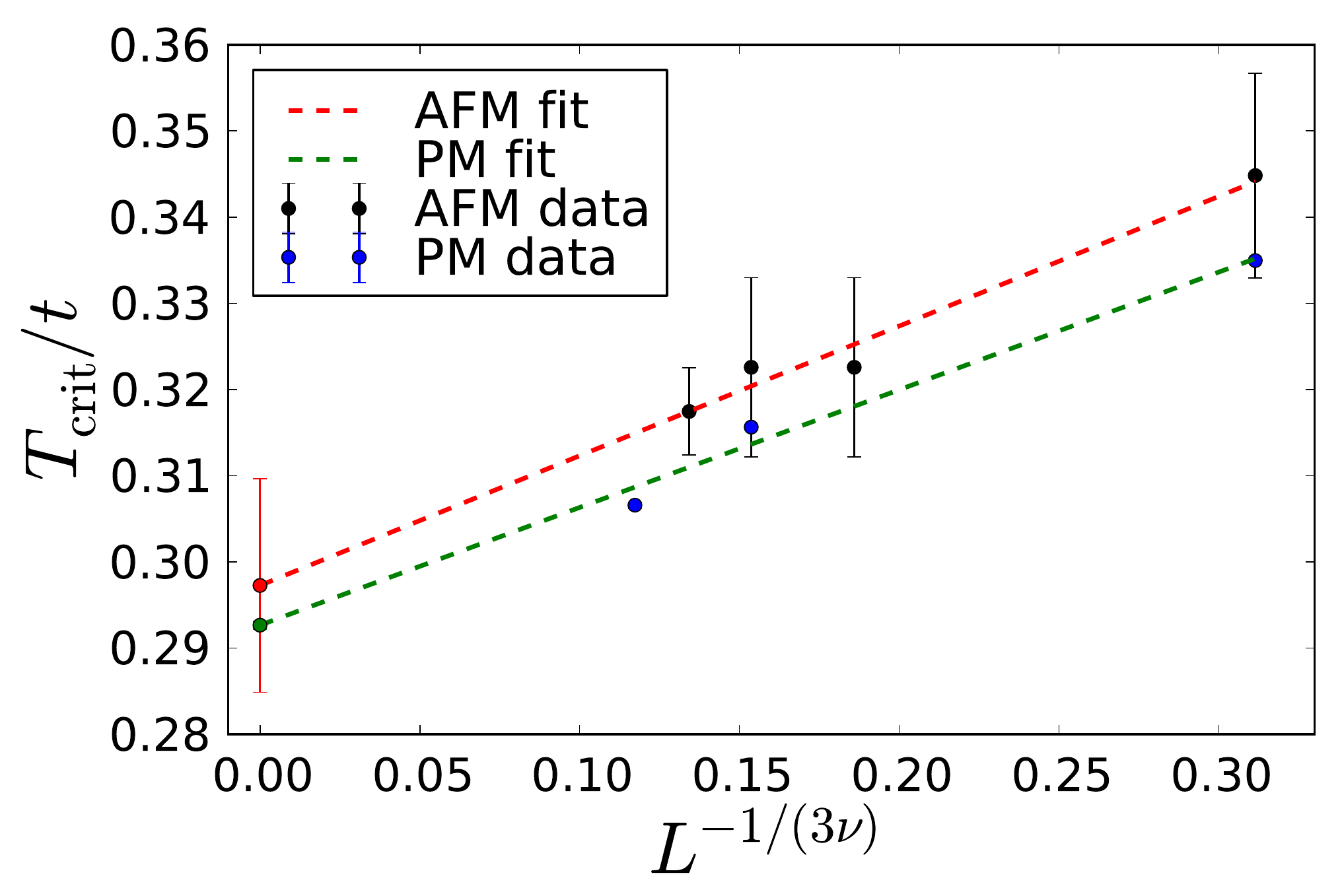}
\caption{Extrapolation of the $T_N$ for the stacked honeycomb for $t/t^\prime=1$, $U/t=6$. The critical exponent $\nu=0.71$ was chosen as that of the the 3D Heisenberg model. The error bars for the AFM data show lower and upper bounds of the transition based on measurement of the staggered magnetization. The PM data points are susceptibility based measurement with error bars smaller than the symbol size.}
\label{fig:Tcrit_extrapolation}
\end{figure}

A comparison of the susceptibility based results measured in the paramagnetic (PM) regime with the results obtained by direct observation of spontaneous magnetization in a formulation with doubled unit cell (with $4$ sites per cell) allowing for antiferromagnetic (AF) ordering are displayed in Fig.~\ref{fig:Tcrit_extrapolation}.

\section{Clusters used in the simulations}
\label{sec:AppendixClusters}
The clusters used in the study are parallelograms defined by cluster basis vectors ${\bf c}_i=(c_{i1},c_{i2},c_{i3})$, $i=1,2,3$ with respect to the lattice basis vectors, i.e. the $i$-th cluster basis vector is in realspace given by $c_{i1}{\bf a}_1+c_{i2}{\bf a}_2+c_{i3}{\bf a}_3$.
A cluster defines a superlattice of which it is the unit cell.

A complete list of clusters used for the EOS calculations is provided in  Tabs.~\ref{tab:ClustersForLayeredSquareLattice},\ref{tab:ClustersForLayeredHoneycombLatticeHalfFilling}, \ref{tab:ClustersForLayeredHoneycombLatticeEOS}, \ref{tab:ClustersForLayeredSquareLatticeTcrit}, and \ref{tab:ClustersForLayeredHoneycombLatticeTcrit}.

\begin{table}
\centering
\begin{tabular}{| c | c | c |}
	\hline
	$t/t^\prime$ & $L$ & cluster basis \\
	\hline
	\multirow{4}{*}{1} & 36 & $(1, 0, 3) , (3, 2, -1) , (2, -2, -2)$ \\
	& 56 & $(1, 2, 3) , (-2, 3, -1) , (-4, -1, 1)$ \\
	& 64 & $(1, 2, 3) , (-2, 4, 2) , (-1, -2, 5)$ \\
	& 74 & $(1, 3, 4) , (3, 4, -1) , (-2, 2, -2)$ \\
	\hline
	\multirow{3}{*}{2} & 8 & $(0,0,1) , ( 2, 2,0) , (-2, 2,0)$ \\
	& 36 & $(0, 0, 2) , (3, 3, 0) , (-3, 3,0)$ \\
	& 102 & $(1, 0, 3) , (5, 3, 0) , (-3, 5, 0)$ \\
	\hline
	\multirow{3}{*}{4} & 8 & $(0, 0,1) , ( 2, 2,0) , ( -2, 2,0)$ \\
	& 18 & $(0, 0, 1) , (3, 3,0) , (0, -3, 3,0)$ \\
	& 116 & $(0, 0,2) , ( 7, 3,0) , (-3, 7,0)$ \\
	\hline
	\multirow{4}{*}{6} & 8 & $(0,0,1) , (2, 2,0) , (-2, 2,0)$ \\
	& 18 & $(0,0,1) , (3, 3,0) , (-3, 3,0)$ \\
	& 34 & $(0,0,1) , (5, 3,0) , (-3, 5,0)$ \\
	& 74 & $(0,0,1) , (7, 5,0) , (-5, 7,0)$ \\
	\hline
	\multirow{3}{*}{8} & 8 & $(0, 0,1) , ( 2, 2,0) , (-2, 2,0)$ \\
	& 18 & $(0, 0,1) , ( 3, 3,0) , (-3, 3,0)$ \\
	& 58 & $(0, 0,1) , (7, 3,0) , (-3, 7,0)$ \\
	\hline
\end{tabular}
\caption{Clusters used for the EOS calculations on the stacked square lattice. For $T/t>4$ we used only the two smallest clusters.}
\label{tab:ClustersForLayeredSquareLattice}
\end{table}

\begin{table}
\centering
\begin{tabular}{| c | c | c | c |}
	\hline
	$t/t^\prime$ & $L$ & temperature & cluster basis\\
	\hline
	\multirow{5}{*}{1} & 6 &  & $(2,-1,0) , (-1,2,0) , (0,0,2)$ \\
	& 18 &  & $(2,2,0) , (0,3,0) , (0,0,3)$ \\
	& 27 &  & $(3,0,0) , (0,3,0) , (0,0,3)$ \\
	& 36 & $T/t<4$ & $(3,0,0) , (0,3,0) , (0,0,4)$ \\
	& 48 & $T/t<4$ & $(4,-2,0) , (-2,4,0) , (0,0,4)$ \\
	\hline
	\multirow{3}{*}{2} & 6 &  & $(2, -1, 0) , (-1, 2, 0) , (0, 0, 1)$ \\
	& 36 &  & $(3, 0, 0) , (0, 3, 0) , (0, 0, 2)$ \\
	& 108 & $T/t<4$ &  $(4, 1, 0) , (-2, 4, 0) , (0, 0, 3)$ \\
	\hline
	\multirow{5}{*}{4, 6} & 3 & $T/t>4$ & $(2, -1, 0) , (-1, 2, 0) , (0, 0, 1) $ \\
	& 9 & & $(3, 0, 0) , (0, 3, 0) , (0, 0, 1)$ \\
	& 66 &  & $(8, -1, 0) , (1, 4, 0) , (0, 0, 2)$ \\
	\hline
	\multirow{3}{*}{8} & 9 &  & $(3, 0, 0) , (0, 3, 0) , (0, 0, 1)$ \\
	& 21 &  & $(5, -4, 0) , (-1, 5, 0) , (0, 0, 1)$ \\
	& 39 & $T/t<4$ & $(7, -5, 0) , (-2, 7, 0) , (0, 0, 1)$ \\
	\hline
\end{tabular}
\caption{Clusters used for the stacked honeycomb lattice to obtain the EOS at half filling.}
\label{tab:ClustersForLayeredHoneycombLatticeHalfFilling}
\end{table}

\begin{table}
\centering
\begin{tabular}{| c | c | c |}
	\hline
	$L$ & temperature & cluster basis\\
	\hline
	3 & $T/t>4$ & $(2, -1, 0) , (-1, 2, 0) , (0, 0, 1) $ \\
	9 &  & $(3, 0, 0) , (0, 3, 0) , (0, 0, 1)$ \\
	12 & $T/t\leq 4$ & $(4, -2, 0) , (-2, 4, 0) , (0, 0, 1)$ \\
	21 & $T/t\leq 4$ & $(5, -4, 0) , (-1, 5, 0) , (0, 0, 1)$ \\
	39 & $T/t\leq 1.5\:^\ast$ & $(7, -5, 0) , (-2, 7, 0) , (0, 0, 1)$ \\
	\hline
\end{tabular}
\caption{Clusters used to obtain the EOS at $(U/t,t/t^\prime)=(6,6)$ and $(4,4)$ for the stacked honeycomb lattice. $^\ast\:$The $39$-cell cluster was not used in case of $(U/t,t/t^\prime)=(4,4)$ for $\mu/t<-2$, in which case three clusters only were used.
}
\label{tab:ClustersForLayeredHoneycombLatticeEOS}
\end{table}

\begin{table}
\centering
\begin{tabular}{| c | c | c |}
	\hline
	$t/t^\prime$ & $L$ & cluster basis \\
	\hline
	\multirow{2}{*}{1} & 36, 56, 64, 74 & same as in Tab.~\ref{tab:ClustersForLayeredSquareLattice} for $t=t^\prime$ \\
	& 128 & $(-2, 4, 4),(2, -4, 4),(1, 6, 1)$ \\
	\hline
	\multirow{4}{*}{2} & 8 & $(1, 0, 1) , (2, 2, 0) , (-2, 2, 0)$ \\
	& 36  & $( 0, 0,2) , ( 3, 3,0) , (-3, 3,0)$ \\
	& 36  & $( 1, 0,2) , ( 3, 3,0) , (-3, 3,0)$ \\
	& 102 & $(1, 0, 3) , (5, 3,0) , (-3, 5,0)$ \\
	\hline
	\multirow{3}{*}{4} &  18 & $( 1, 0,1) , ( 3, 3,0) , (-3, 3,0)$ \\
	%& 36$^{*}$ & $(0, 0,2) , ( 3, 3,0) , (-3, 3,0)$ \\
	& 100 & $(0, 0,2) , ( 5, 5,0) , (-5, 5,0)$ \\
	& 384 & $(1,0,3),(8,8,1),(-8,8,1)$ \\
	\hline
	\multirow{3}{*}{8} & 20 & $(1,0,1),(4,2,0),(-2,4,0)$ \\
	& 50 &  $(1,0,1),(5,5,0),(-5,5,0)$ \\
	& 196 & $(0,0,2),(7,7,0),(-7,7,0)$ \\
	\hline
\end{tabular}
\caption{Bipartite clusters of stacked square lattice used for the $T_N$ estimate. For $t/t^\prime=2$ we used two different $36$-site clusters to minimize potential bias due to cluster choice, as the cluster with basis vector $(0,0,2)$ has every site doubly coupled to its neighbor in the vertical direction due to the periodic boundary conditions.
%The $36$-site cluster for $t=4t^\prime$ marked with $^{*}$ served for check -- it has the same in-plane extent as the $18$-site cluster and as expected, the $T_N$ estimates for both of these clusters are similar; in the extrapolation we assigned an effective size of $18$ sites for both of them.
}
\label{tab:ClustersForLayeredSquareLatticeTcrit}
\end{table}

\begin{table}
\centering
\begin{tabular}{| c | c | c |}
	\hline
	$t/t^\prime$ & $L$ & cluster basis \\
	\hline
	\multirow{3}{*}{1} & 12 & $(0, 0, 4) , (2, -1, 0) , (-1, 2, 0)$ \\
	& 54  & $( 0, 0,6) , ( 3, 0,0) , (0, 3,0)$ \\
	& 96 & $( 0, 0,8) , ( 4, -2,0) , (-2, 4,0)$ \\
	\hline
	\multirow{6}{*}{2} & 18 & $( 0, 0,2) , ( 3, 0,0) , (0, 3,0)$ \\
	& 18 & $(1,0,2),(3,0,0),(0,3,0)$ \\
	& 24$^{*}$ & $(0,0,2),(4,-2,0),(-2,4,0)$ \\
	& 24$^{*}$ & $(1,0,2),(4,-2,0),(-2,4,0)$ \\
	& 36$^{*}$ & $(1,0,2),(4,-2,0),(1,4,0)$ \\
	& 108 & $(0, 0,4) , ( -3, 6,0) , (6, -3,0)$ \\
	\hline
\end{tabular}
\caption{Bipartite clusters of stacked honeycomb lattice used for the $T_N$ estimate. Similarly as for stacked square lattice we used two different $18$-cell clusters to minimize potential bias by particular cluster geometry. The clusters marked with $^{*}$ were used only for $U=6t$.}
\label{tab:ClustersForLayeredHoneycombLatticeTcrit}
\end{table}

\section{Energy estimation and spectral moments}
\label{sec:AppendixSpectralMoments}

The energy per site $e$ was obtained by
\begin{eqnarray}
	e &=& \frac{U}{L\ell}\sum_{{\bf r},\alpha} \langle \hat{n}_{{\bf r}\alpha\uparrow} \hat{n}_{{\bf r}\alpha\downarrow} \rangle 
	      - \frac{1}{L\ell\Omega} \sum_{{\bf K},\sigma} \int_{\textrm{patch}}\mathrm{d}\tilde{\bf k}  \label{eq:energy_per_site} \\
	      && \times \left[ \mathrm{Tr} \left( T_{{\bf K}+\tilde{\bf k}} \right) + \frac{1}{\beta}\sum_{n} \mathrm{Tr} \left( T_{{\bf K}+\tilde{\bf k}} G^\textrm{lat}_\sigma({\bf K}+\tilde{\bf k},i\omega_n) \right) \right], \nonumber
\end{eqnarray}
with $T_{\bf k} = \tilde{T}_{\bf k} - \mu \mathbb{1}$, $\langle \hat{n}_{{\bf r}\alpha\uparrow} \hat{n}_{{\bf r}\alpha\downarrow} \rangle$ measured directly on the cluster, and the lattice Green's function computed as
\begin{equation}
	G^\textrm{lat}_\sigma({\bf K}+\tilde{\bf k},i\omega_n)^{-1} =  G^0_\sigma({\bf K}+\tilde{\bf k},i\omega_n)^{-1}-\Sigma_\sigma({\bf K},i\omega_n).
\end{equation}
The high-frequency tail is added based on spectral moments given below.

The spectral moments of the full Green's function, 
\begin{equation}
	G_\sigma({\bf k},i\omega_n) = \frac{\mathbb{1}_\ell}{i\omega_n} + \frac{C_2^{{\bf k}\sigma}}{(i\omega_n)^2} + \frac{C_3^{{\bf k}\sigma}}{(i\omega_n)^3}+ O\big((i\omega_n)^{-4}\big),
\end{equation}
are used for precise FT from Matsubara representation ($i\omega_n$) to imaginary time representation ($\tau$). In the framework of multisite DCA~in Sec. \ref{sec:AppendixMultisiteDCA} they are given by expressions
\begin{eqnarray}
	\left(C_2^{{\bf k}\sigma}\right)_{\alpha\alpha^\prime} &=& -\tilde{t}_{{\bf k}\alpha\alpha^\prime}+U\delta_{\alpha\alpha^\prime}\left\langle n_{\alpha\bar{\sigma}} \right\rangle , \\
	\left(C_3^{{\bf k}\sigma}\right)_{\alpha\alpha^\prime} &=& \sum_{\gamma} \tilde{t}_{{\bf k}\alpha\gamma} \tilde{t}_{{\bf k}\gamma\alpha^\prime}
	- \tilde{t}_{{\bf k}\alpha\alpha^\prime} U \left\langle n_{\alpha\sigma} + n_{\alpha^\prime\bar{\sigma}}\right\rangle   \nonumber \\
	&& + U^2 \delta_{\alpha\alpha^\prime} \left\langle n_{\alpha\bar{\sigma}} \right\rangle  .
\end{eqnarray}
%using $\tilde{t}_{{\bf k}\alpha\alpha^\prime}=t_{{\bf k}\alpha\alpha^\prime}+\mu\delta_{\alpha\alpha^\prime}$.

\bibliographystyle{apsrev4-1}
\bibliography{layered_honeycomb}

\end{document}